\documentstyle[aps,prb,eqsecnum,multicol,epsf]{revtex}
\begin{document}
\draft
\def\figurename{Fig.}
\renewcommand{\narrowtext}{\begin{multicols}{2} \global\columnwidth20.5pc}
\renewcommand{\widetext}{\end{multicols} \global\columnwidth42.5pc}
\multicolsep = 8pt plus 4pt minus 3pt
\title{Introduction to the Diffusion Monte Carlo Method}

\author{Ioan Kosztin, Byron Faber and Klaus Schulten}
\address{Department of Physics, University of Illinois at Urbana-Champaign,\\
1110 West Green Street, Urbana, Illinois 61801}
\date{August 25, 1995}
\maketitle
\begin{abstract}
  A self--contained and tutorial presentation of the diffusion Monte Carlo
  method for determining the ground state energy and wave function of
  quantum systems is provided.  First, the theoretical basis of the method
  is derived and then a numerical algorithm is formulated. The algorithm is
  applied to determine the ground state of the harmonic oscillator, the
  Morse oscillator, the hydrogen atom, and the electronic ground state of
  the H$^+_2$ ion and of the H$_2$ molecule. A computer program on which the
  sample calculations are based is available upon request.
\end{abstract}

\pacs{E-print: {\tt physics/9702023}}

\narrowtext

\section{Introduction}
\label{sec:intro}

The Schr\"odinger equation provides the accepted description for microscopic
phenomena at non-relativistic energies.  Many molecular and solid state
systems are governed by this equation.  Unfortunately, the Schr\"odinger
equation can be solved analytically only in a few highly idealized cases;
for most realistic systems one needs to resort to numerical descriptions.
In this paper we want to introduce the reader to a relatively recent
numerical method of solving the Schr\"odinger equation, the {\em Diffusion
  Monte Carlo\/} (DMC) method. This method is suitable to describe the
ground state of many quantum systems.

The solution of the time-dependent Schr\"odinger equation can be written as
a linear superposition of stationary states in which the time dependence is
given by a phase factor $\exp\left(-iE_nt/ \hbar\right)$, where $E_n$ is the
$n$-th energy level of the quantum system in question. The energy scale can
be chosen such that all energies are positive. In the DMC method one
actually considers the solution of the Schr\"odinger equation assuming
imaginary time $\tau$, i.e., after replacing the time $t$ by $-i\tau$. The
solution is then given by a sum of transients of the form
$\exp\left(-E_n\tau/\hbar\right)$, $n = 0,1,\ldots$. The DMC method is based
upon the observation that, as a quantum system evolves in imaginary time,
the longest lasting transient corresponds to the ground state with energy
$E_0 < E_n$, $n = 1,2,\ldots$.. Following the evolution of the wave function
in imaginary time long enough one can determine both the ground state energy
$E_0$ and the ground state wave function $\phi_0$ of a quantum system,
regardless of the initial state in which the system had been prepared. The
DMC method provides a practical way of evolving in imaginary time the wave
function of a quantum system and obtaining, ultimately, the ground state
energy and wave function.

The DMC method can be formulated in two different ways. The first one is
based on the similarity between the imaginary time Schr\"odinger equation
and a generalized diffusion equation.  The kinetic (potential) energy term
of the Schr\"odinger equation corresponds to the diffusion ({\em
  source/sink\/} or {\em reaction\/}) term in the generalized diffusion
equation. The diffusion--reaction equation arising can be solved by
employing {\em stochastic calculus} as it was first suggested by Fermi
around 1945\cite{ceperley86,kalos86}. Indeed, the imaginary time
Schr\"odinger equation can be solved by simulating random walks of particles
which are subject to birth/death processes imposed by the source/sink term.
The probability distribution of the random walks is identical to the wave
function. This is possible only for wave functions which are positive
everywhere, a feature, which limits the range of applicability of the DMC
method. Such a formulation of the DMC method was given for the first time by
Anderson\cite{anderson} who used this method to calculate the ground state
energy of small molecules such as H$_3^+$.

A second formulation of the DMC method arises from the Feynman path integral
solution of the time--dependent Schr\"odinger equation. By means of path
integrals the wave function can be expressed as a multi-dimensional integral
which can be evaluated by employing the Monte Carlo method.  Algorithms to
solve the diffusion--reaction equation obeyed by the wave function and
algorithms to evaluate the path integral representation of the wave function
yield essentially one and the same formulation of the DMC method. Which one
of the two formulations of the DMC method one adopts depends on one's
expertise: a formulation of the DMC method based on the diffusion--reaction
equation requires basic knowledge of the theory of stochastic processes; a
path integral formulation obviously requires familiarity with the
corresponding formulation of quantum mechanics.

The purpose of this article is to provide a self-contained and tutorial 
presentation of the path-integral formulation of the DMC method.  We also
present a numerical algorithm and a computer program based on the DMC
method and we apply this program to calculate the ground state energy and
wave function for some sample quantum systems.

The article is organized as follows: The formulation of the DMC method is
presented in Sec.~\ref{sec:dmc}. In Sec.~\ref{sec:sim} a numerical algorithm
for the DMC method is constructed. The results of the DMC simulation for
some simple quantum mechanical systems are presented in Sec.~\ref{sec:app}.
Finally, Sec.~\ref{sec:con} provides suggestions for further numerical
experiments and guides the reader to the literature on the DMC method.

\section{Theory}
\label{sec:dmc}

The theoretical formulation of the DMC method, presented below, 
follows three steps. These steps will be outlined first, to provide the
reader with an overview. 

\noindent {\bf First Step: Imaginary Time Schr\"odinger Equation}. 
In this step, the solution of the time--dependent Schr\"odinger
equation of a quantum system is expressed as a formal series expansion in
terms of the eigenfunctions of the Hamiltonian. One then performs a
transformation from real time $t$ to imaginary time $\tau$, replacing
$t\rightarrow -i\tau$. The solution of the obtained imaginary time
Schr\"odinger equation becomes a series of transients which decay
exponentially as $\tau\rightarrow\infty$.  The longest lasting transient
corresponds to the ground state (i.e., the state with the lowest possible
energy) of the system.

\noindent {\bf Second Step: Path Integral Formulation and Monte Carlo 
  Integration}. In this step, the imaginary time Schr\"odinger
equation is investigated by means of the path integral method.  By using
path integrals the solution of this equation can be reduced to quadrature,
provided that an initial state wave function is given.  Standard Monte Carlo
methods\cite{nr-C} permit one to evaluate numerically the path integral to
any desired accuracy, assuming that the initial state wave function and,
therefore, the ground state wave function as well, is positive definite. In
this case the wave function itself can be interpreted as a probability
density and the ``classical'' Monte Carlo method can be applied.  According
to the general principles of quantum mechanics, only the square of the
absolute value of the wave function has the meaning of a probability
density; the fact that the ground state wave function has to be a positive
definite real quantity imposes severe limitations on the applicability of
the Monte Carlo technique for solving the Schr\"odinger equation.  An
efficient implementation of the standard Monte Carlo algorithm for
calculating the wave function as a large multi-dimensional integral is
realized through an alternation of diffusive displacements and of so-called
birth--death processes applied to a set of imaginary particles, termed
``replicas'', distributed in the configuration space of the system.  The
spatial distribution of these replicas converges to a probability density
which represents the ground state wave function. The diffusive displacements
and birth--death processes can be simulated on a computer using random
number generators.

\noindent {\bf Third Step: Continuous Estimate of the Ground State Energy 
  and Sampling of the Ground State Wave Function}.  In this
step, the ground state energy and the ground state wave function are
actually determined.  As mentioned above, the Monte Carlo method samples the
wave function after each time step. The spatial coordinate distribution of
the replicas involved in the combined diffusion and birth--death processes,
after each (finite) time step, provides an approximation to the wave
function of the system at that given time. The wave function converges in
(imaginary) time towards the (time--independent) ground state wave function,
if and only if the origin of the energy scale is equal to the ground state
energy.  Since the ground state energy is initially unknown, one starts with
a reasonable guess and, after each time step in which a diffusive
displacement and birth--death process is applied to all particles once, one
improves the estimate of the ground state energy.  Ultimately, this
estimate converges towards the desired ground state energy and the
distribution of particles converges to the ground state wave function.

In the following, we shall provide a detailed account of the above
steps.

\subsection{Imaginary Time Schr\"odinger Equation}
\label{sec:tdse}

For simplicity, let us consider a single particle of mass $m$ which moves
along the $x$--axis in a potential $V(x)$. Its wave function $\Psi(x,t)$ is
governed by the time-dependent Schr\"odinger equation\cite{landau3}
\begin{equation}
i\hbar\frac{\partial\Psi}{\partial t} \;=\; \hat H \Psi\;,
  \label{eq:tdse1}
\end{equation}
where the Hamiltonian has the form
\begin{equation}
\hat H \;=\; -\frac{\hbar^2}{2m}\frac{\partial^2}{\partial x^2} \; + \; V(x)\;.
  \label{eq:tdse2}
\end{equation}
Assuming that the potential for $x\, \rightarrow \, \pm \infty$ becomes
infinite, i.e., the particle motion is confined to a finite spatial domain,
the formal solution of (\ref{eq:tdse1}) can be written as a series expansion
in terms of the eigenfunctions of $\hat H$
\begin{equation}
\Psi(x,t) \;=\; \sum_{n=0}^{\infty} c_n \, \phi_n(x) \; e^{-\frac{i}{\hbar}E_n t}\;.
  \label{eq:tdse3}
\end{equation}
The eigenfunctions $\phi_n(x)$, which are square--integrable in the present
case, and the eigenvalues $E_n$ are obtained from the time-independent
Schr\"odinger equation
\begin{equation}
\hat H \phi_n(x) \;=\; E_n \phi_n(x)\; ,
  \label{eq:tdse4}
\end{equation}
subject to the boundary conditions $\lim_{x\rightarrow \pm \infty} \phi_n(x)\, = \, 0$.  
We label the energy eigenstates by $n \;=\; 0,1,2,\ldots\;$ and order the
energies 
\begin{equation}
E_0 < E_1 \leq E_2 \leq\ldots
  \label{eq:tdse5}
\end{equation}
The eigenfunctions $\phi_n(x)$ are assumed to be orthonormal and real, i.e., 
\begin{equation}
\int_{-\infty}^{\infty} dx\phi_n(x) \phi_m(x) \;=\; \delta_{nm}\;.
  \label{eq:tdse6}
\end{equation}
The expansion coefficients $c_n$ in (\ref{eq:tdse3}) are then 
\begin{equation}
c_n \;=\; \int_{-\infty}^{\infty} dx \phi_n(x) \Psi(x,0) \;,
\qquad n\;=\;0,1,2,\ldots \; ,
  \label{eq:tdse7}
\end{equation}
i.e., they describe  the overlap of the initial state  $\Psi(x,0)$, else
assumed real, with the eigenfunctions
$\phi_n(x)$ in (\ref{eq:tdse4}). 

\noindent {\bf Shift of Energy Scale}.  We perform now a trivial, but 
methodologically crucial shift of the energy scale introducing the
replacements $V(x)\, \rightarrow V(x)\, - \, E_R$ and $E_n\, \rightarrow
E_n\, - \, E_R$.  This leads to the Schr\"odinger equation
\begin{equation}
  i \hbar\frac{\partial\Psi}{\partial\tau} \;=\; -\,
  \frac{\hbar^2}{2m}\frac{\partial^2\Psi}{\partial x^2} \, + \, \left[V(x) -
    E_R\right]\Psi\;,
  \label{eq:tdse8a}
\end{equation}
and to  the expansion  
\begin{equation}
\Psi(x,t) \;=\; \sum_{n=0}^{\infty} c_n\,  \phi_n(x)\;  e^{- i \frac{E_n - E_R}{\hbar}t}\;.
\label{eq:tdse9a}
\end{equation}

\noindent {\bf Wick Rotation of Time}.  
Now let us perform a transformation from real time to imaginary time (also
known as Wick rotation) by introducing the new variable $\tau \;=\; it$.
The Schr\"odinger equation (\ref{eq:tdse8a}) becomes
\begin{equation}
\hbar\frac{\partial\Psi}{\partial\tau}  \;=\;
\frac{\hbar^2}{2m}\frac{\partial^2\Psi}{\partial x^2} - \left[V(x) - E_R\right]\Psi\;,
  \label{eq:tdse8}
\end{equation}
and the expansion  (\ref{eq:tdse9a})  reads 
\begin{equation}
\Psi(x,\tau) \;=\; \sum_{n=0}^{\infty} c_n \;  \phi_n(x) \;  e^{-\frac{E_n - E_R}{\hbar}\tau}\;.
  \label{eq:tdse9}
\end{equation}

Noting the energy ordering  (\ref{eq:tdse5}), one can infer from
(\ref{eq:tdse9}) the following asymptotic behavior for $\tau\rightarrow\infty$
\begin{itemize}
\item [(i)] if $E_R > E_0, \quad
  \lim_{\tau\rightarrow\infty}\Psi(x,\tau) = \infty$, the
  wave function diverges exponentially fast;
\item [(ii)] if $E_R < E_0, \quad
  \lim_{\tau\rightarrow\infty}\Psi(x,\tau) = 0$, the wave
  function vanishes exponentially fast;
\item [(iii)] if $E_R = E_0, \quad
  \lim_{\tau\rightarrow\infty}\Psi(x,\tau) = c_0 \phi_0(x)$,
  the wave function converges, up to a constant factor $c_0$ defined through
  (\ref{eq:tdse7}), to the ground state wave function. 
\end{itemize}
This behavior provides the basis of the DMC method.  For $E_R = E_0$, the
function $\Psi(x,\tau)$ converges to the ground state wave function
$\phi_0(x)$ regardless of the choice of the initial wave function
$\Psi(x,0)$, as long as there is a numerically significant overlap between
$\Psi(x,0)$ and $\phi_0(x)$, i.e., as long as $c_0$ is not too small.  The
ground state wave function, for a single particle, has no nodes (in case of
many fermion systems this might not be true) and one can always fulfill the
requirement of non--vanishing $c_0$ by choosing a positive definite initial
wave function centered in a region of space where $\phi_0(x)$ is
sufficiently large.

We now seek a practical way to integrate equation (\ref{eq:tdse8}) for an
arbitrary reference energy $E_R$ and initial wave function $\Psi(x,0)$. We
shall accomplish this by using the path integral formalism.

\subsection{Path Integral Formalism}
\label{sec:pif}

\noindent The solution of the imaginary time Schr\"odinger equation
(\ref{eq:tdse8}) can be written
\begin{equation}
\Psi(x,\tau) \; =\;  \int_{-\infty}^{\infty}\!\!\!dx_0\:
K\!\left(x,\tau|x_0,0\right) \Psi\left(x_0,0\right)\;,
  \label{eq:pif1}
\end{equation}
where the propagator $K\!\left(x,\tau|x_0,0\right)$ is expressed in terms of the well-known
path integral \cite{feynman}, modified by the replacement $t\, = \, -i\tau$
\begin{eqnarray}
& & K\!\left(x,\tau|x_0,0\right)  = \lim_{N\rightarrow\infty}
\int_{-\infty}^{\infty}\!\!\!dx_1 \ldots
\int_{-\infty}^{\infty}\!\!\!dx_{N-1}
\left(\frac{m}{2\pi\hbar\Delta\tau}\right)^{\frac{N}{2}}\!\!\!\times 
\nonumber \\
& & \label{eq:pif2}
\\
& & \exp\left\{-\frac{\Delta\tau}{\hbar}\sum_{j=1}^{N}\left[\frac{m}{2\Delta\tau^2}\left(x_j-x_{j-1}\right)^2+V\left(x_j\right)-E_R\right]\right\}\;.\nonumber
\end{eqnarray}
Here $\Delta\tau = \tau/N$ is a small time step.  One sets  $x_N\;\equiv\;x$.
The wave function $\Psi(x,\tau)$ can be written in the form 
\[
\Psi(x,\tau) \;=\; 
\lim_{N\rightarrow\infty}
\int_{-\infty}^{\infty}\left(\prod_{j=0}^{N-1} dx_j\right)\:
\prod_{n=1}^{N} W\left(x_n \right)\times
\]
\begin{equation}
\times P\left(x_n,x_{n-1}\right)
\Psi\left(x_0,0\right)\;,
\label{eq:pif3} 
\end{equation}
where we have defined
\begin{equation}
P\left(x_n,x_{n-1}\right) \;\equiv\; 
\left(\frac{m}{2\pi\hbar\Delta\tau}\right)^{\frac{1}{2}}
\exp\left[-\frac{m\left(x_n-x_{n-1}\right)^2}{2\hbar\Delta\tau}\right]\;, 
  \label{eq:pif4}
\end{equation}
and
\begin{equation}
W\left(x_n\right)\;\equiv\; 
\exp\left[-\frac{\left[V\left(x_n\right)-E_R\right]\Delta\tau}{\hbar}\right]\;.
  \label{eq:pif5}
\end{equation}
The function $P\left(x_n,x_{n-1}\right)$ is related to the kinetic energy
term in (\ref{eq:tdse2}). This function can be thought of as a Gaussian
probability density for the random variable $x_n$ with mean equal to
$x_{n-1}$ and variance
\begin{equation}
  \label{eq:pif5d}
  \sigma\;=\;\sqrt{\hbar\Delta\tau/m}\;.
\end{equation}
The so-called {\em weight function\/} $W\left(x_n\right)$ depends on both
the potential energy in (\ref{eq:tdse2}) and the reference energy $E_R$. The
main difference between the functions $P$ and $W$ is that the former can be
interpreted as a probability density since
\begin{equation}
  \label{eq:pif5e}
    \int_{-\infty}^{\infty}\!\!dy P(x,y)\;=\;1\;,
\end{equation}
while the latter can not.

The path integral (\ref{eq:pif3}) can be evaluated analytically only for
particular forms of the potential energy $V(x)$\cite{khandekar93}.
Fortunately, by choosing $N$ sufficiently large, one can evaluate
(\ref{eq:pif3}) numerically to any desired accuracy. However, since a
suitable $N$ is necessarily a large number, the standard algorithms of
numerical integration cannot be employed directly\cite{koonin}, instead, one
uses the so-called Monte Carlo method\cite{nr-C}.  According to this method
any (convergent) $N$--dimensional integral of the form
\begin{equation}
I \;=\;
\int_{-\infty}^{\infty}\left(\prod_{j=0}^{N-1} dx_j\right)\:
f\left(x_0,\ldots,x_{N-1}\right){\cal P}\left(x_0,\ldots,x_{N-1}\right)\;,
  \label{eq:pif6}
\end{equation}
where ${\cal P}$ is a probability density, i.e.,
\[
{\cal P}\left(x_0,\ldots,x_{N-1}\right) \;\geq\; 0\;,\qquad\text{and}
\]
\begin{equation}
\int_{-\infty}^{\infty}\left(\prod_{j=0}^{N-1} dx_j\right)\:
  {\cal P}\left(x_0,\ldots,x_{N-1}\right) \;=\;1\;,
  \label{eq:pif7}
\end{equation}
can be approximated by the expression
\begin{equation}
{\cal I} \;=\; \frac{1}{{\cal N}}\sum_{\stackrel{i=1}{x^{(i)}\in{\cal P}}}^{\cal N}
f\left(x_0^{(i)},\ldots,x_{N-1}^{(i)}\right)\;.
  \label{eq:pif9}
\end{equation}
Here the notation $x^{(i)}\in{\cal P}$ means that the numbers
$x_j^{(i)}$, $i=1,2,\ldots,{\cal N}$; $j=0,\ldots,N-1$, are
selected randomly with the probability density ${\cal P}$. The larger
${\cal N}$, the better is the approximation $I \approx {\cal I}$. In
fact, according to the {\em central limit theorem\/}\cite{nr-C}, the
values of ${\cal I}$ obtained as a result of different simulations are
distributed normally, i.e., according to a Gaussian distribution,
around the exact value $I$, the standard deviation
being proportional to $1/\sqrt{\cal N}$.  

In order to evaluate $\Psi(x,\tau)$ in (\ref{eq:pif3}), for given $x$,
$\tau$ and $N$, one defines
\begin{equation}
  \label{eq:pif5a}
  {\cal P}\left(x_0,\ldots,x_{N-1}\right)\;=\;
  \prod_{n=1}^{N} P\left(x_n,x_{n-1}\right)\;,
\end{equation}
and
\begin{equation}
  \label{eq:pif5b}
  f\left(x_0,\ldots,x_{N-1}\right)\;=\;
  \Psi\left(x_0,0\right)\prod_{n=1}^{N} W\left(x_n \right)\;,
\end{equation}
such that one can apply (\ref{eq:pif6}) and (\ref{eq:pif9}). For this
purpose we note that due to
\begin{equation}
  \label{eq:pif5c}
  \int_{-\infty}^{\infty}\!\!dy P(x,y) P(y,z)\;=\;P(x,z)\;, 
\end{equation}
and (\ref{eq:pif5e}) the probability distribution (\ref{eq:pif5a}) does
indeed obey the property (\ref{eq:pif7}). The expression (\ref{eq:pif9}) can
now be invoked to evaluate $\Psi(x,\tau)$ by means the path integral
(\ref{eq:pif3}). This requires the generation of sets of coordinate vectors
$x^{(i)}\in{\cal P}$,
$x^{(i)}=\left(x_0^{(i)},x_1^{(i)},\ldots,x_N^{(i)}\right)$ for
$i=1,2,\ldots,{\cal N}$, where $x_N^{(i)}=x$. 

In order to obtain vectors $x^{(i)}$ which sample the probability density
${\cal P}$ one proceeds as follows: In a first step one generates, for a
fixed $x=x_N^{(i)}$, a Gaussian random number $x_{N-1}^{(i)}$ with mean
value $x_N^{(i)}$ (i.e., a Gaussian random number distributed about
$x_N^{(i)}$) and variance $\sigma$, according to the probability density
$P\left(x_N^{(i)},x_{N-1}^{(i)}\right)$ given by (\ref{eq:pif4}). In a
second step, a Gaussian random number $x_{N-2}^{(i)}$, with mean
$x_{N-1}^{(i)}$ and variance $\sigma$, is generated according to
$P\left(x_{N-1}^{(i)},x_{N-2}^{(i)}\right)$. The steps are continued to
produce random numbers $x_n^{(i)}$ until one reaches $x_0^{(i)}$. 
Two consecutive random numbers $x_n^{(i)}$ and $x_{n-1}^{(i)}$ are
related through the equation
\begin{equation}
  \label{eq:pif8a}
  x_n^{(i)}\;=\;x_{n-1}^{(i)} + \sigma \rho_n^{(i)}\;,
\end{equation}
where $\sigma$ is given by (\ref{eq:pif5d}) and $\rho_n^{(i)}$ is a Gaussian
random number with zero mean and a variance equal to one. The
$\rho_n^{(i)}$'s can be generated numerically by means of algorithms
referred to as random number generators.  One can check, using
(\ref{eq:pif8a}) and (\ref{eq:pif4}), that the mean and the variance of
$x_n^{(i)}$ $\left(x_{n-1}^{(i)}\right)$ are equal to $x_{n-1}^{(i)}$
$\left(x_n^{(i)}\right)$ and $\sigma$, respectively.  Therefore, the
coordinate vectors $\left\{x_{N-1}^{(i)},\ldots,x_0^{(i)}\right\}$, obtained
through equation (\ref{eq:pif8a}) for $i=1,2,\ldots,{\cal N}$, are
distributed according to the probability density (\ref{eq:pif5a}). We note
in passing that the sequence of positions $x_n$ given by (\ref{eq:pif8a})
defines a {\em stochastic\/} process, namely the well known Brownian
diffusion process.

Repeating ${\cal N}$ times the sampling of ${\cal P}$ the wave function
$\Psi(x,\tau)$ can be determined according to (\ref{eq:pif9}) and
(\ref{eq:pif3}). Unfortunately, the algorithm outlined is impractical since
it provides only a route to calculate $\Psi(x,\tau)$ for a chosen time
$\tau$, but no systematic method to obtain the ground state energy and wave
function, which requires a description for $\tau\rightarrow\infty$.

Fortunately, the above so-called Monte Carlo algorithm can be improved and
used to determine {\em simultaneously\/} both $E_0$ and $\phi_0$. The basic
idea is to consider the wave function itself a probability density.  This
implies that the wave function should be a positive definite function, a
constraint which limits the applicability of the suggested method. By
sampling the initial wave function $\Psi(x,0)$ at ${\cal N}_0$ points one
generates as many Gaussian random walks which evolve in time according to
(\ref{eq:pif5a}) or, equivalently, according to (\ref{eq:pif8a}); instead of
tracing the motion of each random walk separately, one rather follows the
motion of a whole ensemble of random walks simultaneously.  The advantage of
this procedure is that one can sample the wave function of the system,
through the actual position of the random walks and the products of the
weights $W$ along the corresponding trajectories, after each time step
$\Delta\tau$. This procedure, as we explain below, also provides the
possibility to readjust the value of $E_R$ after each time step and to
follow the time evolution of the system for as many time steps as are needed
to converge to the ground state wave function and energy .

The procedure, the so-called DMC method, interprets the integrand in
(\ref{eq:pif3}), i.e.,
\[
\underbrace{W\left(x_N \right)}_{{\rm process}~ 2N} \; \underbrace{P\left(x_N,x_{N-1} \right)}_{{\rm process}~ 2N-1} \quad  \cdots  \quad 
\underbrace{ W\left(x_2 \right)}_{{\rm process}~ 4}\;  \underbrace{
  P\left(x_2,x_{1}\right)}_{{\rm process}~ 3}\times
\]
\begin{equation}
\times\underbrace{W\left(x_1 \right)}_{{\rm process}~ 2} \; 
\underbrace{P\left(x_1,x_{0}\right)}_{{\rm process}~ 1} \; 
\underbrace{ \Psi\left(x_0,0\right)}_{{\rm process}~ 0}\;,
\label{eq:pif9aa}
\end{equation}
as a product of probabilities and weights to be modeled by a series of
sequential stochastic processes $0, 1, 2, \ldots 2N$.  We will explain now
how these processes are described numerically.

\noindent {\bf Initial State}. 
The 0-th process describes ``particles'' (random walks) distributed
according to the initial wave function $\Psi\left(x_0,0\right)$, which is
typically chosen as a {\em Dirac} $\delta$--function
\begin{equation}
  \Psi\left(x_0,0\right) \; = \; \delta(x - x_0)\;,
\label{eq:pif9ab}
\end{equation}
where $x_0$ is located in an area where the ground state of the quantum
system is expected to be large.  The initial distribution (\ref{eq:pif9ab})
is obtained by simply placing all particles initially at position $x_0$.

\noindent {\bf Diffusive Displacement}.  As we explained earlier,  
the successive positions $x_{n-1}, \, x_{n}$ in (\ref{eq:pif9aa}) can be
generated through (\ref{eq:pif8a}). The Monte Carlo algorithm produces then
the positions $x_1\, = \, x_0\, + \, \sigma\rho_1$, $x_2\, = \, x_1\, + \,
\sigma\rho_2$, etc by generating the series of
random numbers $\rho_n, \, n\, = \, 1, 2, \ldots$.

\noindent {\bf Birth--Death Process}. Instead of accumulating the product
of the weight factors $W$ for each particle, it is more efficient
numerically to replicate (see below) the particles after each time step with
a probability proportional to $W(x_n)$. In this way, after each time step
$\Delta\tau$, the (unnormalized) wave function is given by a histogram of
the spatial distribution of the particles. The calculation of the wave
function $\Psi(x,\tau)$ can be regarded as a simulated diffusion--reaction
process of some imaginary particles.

In the replication process each particle is replaced by a number of

\begin{equation}
m_n \; = \; {\rm min} \left[\, {\rm int}[ W(x_n) \, + \, u\right], \, 3]
\label{eq:pif9af}
\end{equation}
particles, where ${\rm int}[ x]$ denotes the integer part of $x$ and where
$u$ represents a random number uniformly distributed in the interval
$[0,1]$.  In case $m_n \, = \, 0$ the particle is deleted and one stops the
diffusion process; this is referred to as a ``death'' of a particle.  In
case of $m_n \, = \, 1$ the particle is unaffected and one continues with
the next diffusion step.  In case $m_n \, = \, 2, 3 $ one continues with the
next diffusion step, but begins also a new series (in case of $m_n\, = \, 3$
two new series) of diffusive displacements starting at the present location
$x_n$. This latter case is referred to as the ``birth'' of a particle (of
two particles for $m_n=3$). From (\ref{eq:pif9af}) one can see that at most
two new particles can be generated whereas one would expect that for $ {\rm
  int}[ W(x_n) \, + \, u ] \, \ge \, 4$ three or more new series would be
started. This limitation on the birth rate of the particles is necessary in
order to avoid numerical instabilities, especially at the beginning of the
Monte Carlo simulation, when $E_R$ may differ significantly from $E_0$.  The
error resulting from the limitation $m_n\le 3$ is expected to be small since
for sufficiently small $\Delta\tau$ holds
\begin{equation}
W(x) \; \approx \; 1 \; - \; {V(x)\, - \, E_R\over \hbar}\, \Delta \tau
\label{eq:pif9afa}
\end{equation}
which, evidently, assumes values around unity.

\subsection{The Diffusion Monte Carlo Method}
\label{sec:dmc1}

We want to summarize now  the algorithmic steps actually taken in a straightforward application of the DMC method.  To realize the suggested algorithm one starts  with 
${\cal N}_0$ ``particles'', referred to as ``replicas'',  which are placed according to a distribution $\Psi\left(x_0,0\right)$.      As pointed out above, the actual numbers of replicas will vary as replicas ``die'' and new ones are ``born''.    The replicas are characterized through a position $x_n^{(j)}$ where the  suffix $n$ counts the diffusive displacements and where $(j)$
counts the replicas.  The number of replicas after $n$ diffusive displacements will be denoted by ${\cal N}_n$.  In the initial step one samples ${\cal N}_0$ replicas assigning positions  $x_0^{(i)}$, $\left(i=1,\ldots,{\cal N}_0\right)$ according to the  distribution $\Psi\left(x_0,0\right)$.  As mentioned, we actually choose all replicas to begin at the same point $x_0$.

Rather than following the fate of each replica and of its descendants through many diffusive displacements, one follows all replicas simultaneously.  Accordingly, one determines  the positions $x_1^{(j)}$ following equation (\ref{eq:pif8a}), i.e., one sets 
\begin{equation}
x_{1}^{(j)} \; = \; x_{0}^{(j)} \; + \; \sqrt{\hbar \Delta\tau / m} \cdot \rho_1^{(j)}
\label{eq:pif9ag}
\end{equation}
for replicas $j\, = \, 1, 2, \ldots {\cal N}_0$ generating appropriate
random numbers $\rho_1^{(j)}$.  This can be regarded as a one--step
diffusion process of the system of replicas.  Once the new positions
(\ref{eq:pif9ag}) have been determined, one evaluates the weight
$W(x_1^{(i)})$ given by (\ref{eq:pif5}) and, according to (\ref{eq:pif9af}),
one determines the set of integers $m_1^{(j)}$ for $j\, = \, 1, 2, \ldots
{\cal N}_0$.  Replicas $j$ with $m_1^{(j)}\, = \, 0$ are terminated,
replicas $m_1^{(j)}\, = \, 1, 2, 3$ are left unaffected, except, that in
case $m_1^{(j)}\, = \, 2, 3 $ one or two new replicas $j^\prime$ are added
to the system and their position is set to $x_{1}^{(j^\prime)}\, = \,
x_{1}^{(j)}$.  The number of replicas is counted and, thus, ${\cal N}_1$ is
determined.  During the combined diffusion and birth--death process the
distribution of replicas changes in such a way that the corresponding
coordinates $x_1^{(j)}$ will now be distributed according to a probability
density which will be identified with $\Psi(x,\Delta\tau)$.

\noindent {\bf Adaptation of the Reference Energy $\bbox{E_R}$}.  As a result of the birth--death process the total
number ${\cal N}_1$ of replicas changes from its original ${\cal N}_0$ value.  As discussed above, for $E_R$ values less then the ground state energy the distribution $\Psi(x, \tau)$ decays asymptotically to zero, i.e., the replicas eventually all die; for $E_R$ values larger than the ground state energy, the distribution will increase indefinitely, i.e., the number ${\cal N}_n$ will exceed all bounds.  Only for $E_R\, = \, E_0$ can one expect a stable asymptotic distribution such that the  number of replicas  fluctuates around an average value ${\cal N}_0$.  The spatial distribution and  increases/decreases of the number of replicas allow one to  adjust the value of $E_R$ 
 as to keep the total number of replicas approximately constant.
To this end, one proceeds from (\ref{eq:pif9afa}).  Averaged over all replicas this equation reads
\begin{equation}
\langle W\rangle_1 \; \approx \; 1 \; - \; {\langle V\rangle_1 \, - \, E_R\over \hbar} \, \Delta \tau 
\label{eq:dmcm4a}
\end{equation}
where the average potential energy is 
\begin{equation}
\langle V\rangle_1 \;=\; {1\over {\cal N}_1} \sum_{j=1}^{{\cal N}_1} V(x_1^{(j)})
\label{eq:dmcm5}
\end{equation}
The reader should note that this average varies with the number of diffusion steps taken for all replicas;  in the present case we consider the average after the first diffusive displacement.  
One would like the value of $\langle W\rangle$ to be eventually always unity such that the number of replicas remains constant.  This leads to the proper choice of $E_R$
\begin{equation}
E_R^{(1)}\; = \; \langle V \rangle_1
\label{eq:dmcm4b}
\end{equation}
However, if the distribution of replicas deviates too strongly from $\phi_0(x)$ the number of replicas can experience strong changes which need to be repaired by `overcompensating' through a suitable choice of $E_R$.  In fact, in case ${\cal N}_1/ {\cal N}_0 \, < \,  1$ one would like to increase subsequently the number of replicas in order to restore the initial number of replicas and, hence, choose an $E_R$ value larger than (\ref{eq:dmcm4b});  in case  ${\cal N}_1/ {\cal N}_0 \, > \,  1$ one would like to decrease subsequently the number of replicas and, hence, choose an $E_R$ value smaller  than (\ref{eq:dmcm4b}).  A suitable choice is 
\begin{equation}
E_R^{(2)} \;=\; \langle V\rangle_1 + \frac{\hbar}{\Delta\tau} \left(1-\frac{{\cal N}_1}{{\cal N}_0}\right)\;.
\label{eq:dmcm4}
\end{equation}
In order to justify that (\ref{eq:dmcm4}) serves our purpose one notes
that ${\rm exp}[-\Delta \tau (\langle V\rangle - E_R)/\hbar]$, for
sufficiently small $\Delta \tau$, is $1 -\Delta \tau (\langle V\rangle
- E_R)/\hbar$.  In the subsequent birth--death process this would lead
to an ${\cal N}_1$ value (recall that during the diffusion process the
number of replicas does not change) related to $E_R$ by
\begin{equation}
E_R \;=\; \langle V\rangle_1 - \frac{\hbar}{\Delta\tau}
\left(1-\frac{{\cal N}_1}{{\cal N}_0}\right)\;. 
\label{eq:dmcm4c}
\end{equation}
However, one would like the total number of replicas during the next
birth--death process to return to the initial value ${\cal N}_0$ and $E_R$
to be given by an expression similar to (\ref{eq:dmcm4b}).  This can be
achieved by adding to both sides of equation (\ref{eq:dmcm4c}) the same
quantity $\hbar\left(1-{\cal N}_1/{\cal N}_0\right)/\Delta\tau$. Hence, the
redefined value for the reference energy is
\begin{equation}
E_R^{(2)} \;=\; E_R^{(1)} + \frac{\hbar}{\Delta\tau}\left(1- \frac{{\cal
    N}_1}{{\cal N}_0}\right)\;,
  \label{eq:dmcm4d}
\end{equation}
which, by taking (\ref{eq:dmcm4b}) into account, is identical to
(\ref{eq:dmcm4}). 

Equation (\ref{eq:dmcm4}) should be regarded as an empirical result rather
than an exact one. In fact, any expression of the form $E_R = \langle
V\rangle + \alpha\left(1- {\cal N}/{\cal N}_0 \right)$, with arbitrary
positive $\alpha$, can be used equally well in place of
(\ref{eq:dmcm4})\cite{suhm91}. Usually, the actual value of the ``feedback''
parameter $\alpha$ is chosen empirically for each individual problem so as to
reduce as much as possible the statistical fluctuations in ${\cal N}_0$ and,
at the same time, to diminish unwanted correlations between the
successive generation of replicas.  Equation (\ref{eq:dmcm4}) suggests that
a good starting value for $\alpha$ is $\hbar/\Delta\tau$, a value used 
in our DMC program.

\noindent {\bf 2nd, 3rd, \ldots Displacements}. The diffusive
displacements, birth--death processes and estimates for new $E_R$
values are repeated until, after a sufficiently large number of steps,
the energy $E_R$ and the distribution of replicas becomes stationary.
The ground state energy is then
\begin{equation}
E_0 \;=\; \lim_{n->\infty} \langle V \rangle_n
\label{eq:dmcm8}
\end{equation}
since ${\cal N}_n / {\cal N}_{n-1}\, \rightarrow \, 1$.  The distribution of
replicas provides the ground state function $\phi_0(x)$.

\subsection{Systems with Several Degrees of Freedom}
\label{sec:sdf}

The DMC method described above is valid for a
quantum system with one degree of freedom $x$. However, the
method can be easily extended to quantum systems with several degrees of
freedom.   Such systems arise, for example, in case of  a particle moving in two or three spatial dimensions or in case of several interacting particles. For these two cases
the DMC method can be readily generalized.

\noindent {\bf Particle in $\bbox{d}$ Dimensions}.
The Hamiltonian of a particle of mass $m$ moving in a
potential $V\left(x_1, \ldots x_d \right)$ can be written
\begin{equation}
\hat H \;=\; -\frac{\hbar^2}{2m}\sum_{\alpha=1}^d 
\frac{\partial^2}{\partial x^2_{\alpha}} + V(x_1, \ldots, x_d )\;,
  \label{eq:sdf1}
\end{equation}
where $x_{\alpha}$, $\alpha = 1,...,d$, denotes the Cartesian
coordinates of the particle and $d\:(=2,3)$ represents the spatial 
dimension. For the Hamiltonian (\ref{eq:sdf1}), the
DMC algorithm can be devised in a similar way as for
(\ref{eq:tdse2}). The only difference to the one degree of freedom
case is that during each time step $\Delta\tau$ one needs to execute
$d$  random walks for each replica. Indeed, the kinetic energy term in
(\ref{eq:sdf1}) can be formally regarded as the sum of kinetic
energies of $d$ ``particles'' having the same mass $m$ and moving
along the $x_{\alpha}$, $\alpha = 1,...,d$, directions. Consequently,
the diffusive displacements are governed by the distribution 
\begin{eqnarray}
& P\left(x_{n,1},x_{n-1,1}; \dots\;  x_{n,d},x_{n-1,d}\right) \;=\;  & 
 \label{eq:sdf2} \\
& \prod_{\alpha=1}^d
\left(\frac{m}{2\pi\hbar\Delta\tau}\right)^\frac{1}{2}
\exp\left[-\frac{m\left(x_{n,\alpha}-x_{n-1,\alpha}\right)^2}{2\hbar\Delta\tau}\right]\;.
&
\nonumber 
\end{eqnarray}
The product of probabilities can be described through independent random processes, i.e., one can reproduce the probability (\ref{eq:sdf2}) through $d$ independent diffusive displacements applied for each replica to the $d$ spatial directions.

\noindent {\bf $\bbox{S}$ Particles}.  In the case of $S$ interacting particles, which move in
$d$ spatial dimensions, the most general form of the Hamiltonian is
given by (assuming that no internal, e.g., spin, degrees of freedom
are involved)
\begin{equation}
\hat H \;=\; \sum_{j=1}^{S} \left[ \, - \, \frac{\hbar^2}{2m_j} \sum_{\alpha=1}^d
\frac{\partial^2}{\partial x_{j\alpha}^2} \; + \;  V\left(\{x_{j\alpha}\}\right)\, \right] \;,
\label{eq:sdf3}
\end{equation}
where $V\left(\{x_{j\alpha}\}\right)$ accounts for  both an  interaction
between particles and for an  interaction due to an external field; $\{x_{j\alpha}\}$ denotes the dependence on the coordinates of all particles.  By rescaling the
coordinates in (\ref{eq:sdf3}) 
\begin{equation}
x_{j\alpha} \;=\; \sqrt{\frac{m}{m_j}} x'_{j\alpha}\;, \qquad
j=1,\ldots,S\;, \quad \alpha=1,\ldots,d 
  \label{eq:sdf4}
\end{equation}
where $m$ is an arbitrary mass, one can make the Hamiltonian (\ref{eq:sdf3})  look
formally the same as the Hamiltonian (\ref{eq:sdf1})  describing a single particle of mass $m$
which moves in $d' = S\times d$ spatial dimensions. Hence, the
generalization of the DMC algorithm for this case is
again apparent.

\noindent {\bf Sign Problem}. The case in which  (\ref{eq:sdf3}) actually describes a system of {\em  identical\/} particles  cannot be 
treated  like that of a single particle  moving in $d'
= S\times d$ spatial dimensions. For such systems a prescribed {\sl boson\/} or {\sl fermion\/} symmetry of the wave
function with respect to an interchange of particles must be obeyed. For  bosons (particles with
integer spin) the total wave function (i.e., the product of the
orbital and the spin wave functions) is symmetric with respect to any
permutation of the particles while for  fermions (particles
with half--integer spin) the total wave function is antisymmetric with
respect to such  permutation. This constraint determines
the symmetry of the orbital part of the ground state wave function for
fermionic systems (but not for bosonic ones): in many cases of
interest, the (orbital) ground state wave function of fermionic
systems will have nodes, i.e.,  regions with different signs,  which make
the DMC method, as presented here, inapplicable. We
shall not address this issue in further detail and shall consider only
cases where the so-called ``sign problem'' of the ground state wave
function does not arise.

To summarize, a DMC algorithm for a system with one degree of freedom can be
adopted to a system of $S$ interacting particles moving in $d$ spatial
dimensions with an effective dimension of $d' = S\times d$. Exactly this
feature of the DMC method makes it so attractive for the evaluation of the
ground state of a quantum system.  However, in the case of identical
fermions, one needs to obey the actual symmetry of the ground state wave
function and the method often is not applicable.

\section{Algorithm}
\label{sec:sim}

Our goal is to provide an algorithm for the DMC method presented in the
previous section and to apply this algorithm to obtain the ground state
energy and wave function for sample quantum systems.  Some of the examples
chosen below, e.g., the harmonic oscillator, have an analytical solution
and, therefore, allow one to test the diffusion Monte Carlo method. Other
examples, e.g., the hydrogen molecule, can not be solved analytically and,
hence, the DMC method provides a convenient way of solving the problem. The
obtained results turn out to be in good agreement with results obtained by
means of other numerical methods\cite{herzberg}.

\subsection{Dimensionless Units}
\label{sec:du}

In order to implement the DMC method into a numerical algorithm, one needs
to rewrite all the relevant equations in dimensionless units.  One can go
from conventional (e.g., SI) units to dimensionless units by explicitly
writing each physical quantity as its magnitude times the corresponding
unit. In mechanics the unit of any physical quantity can be expressed as a
proper combination of three independent units, such as, $L$, $T$ and ${\cal
  E}$, which denote the unit of length, time and energy, respectively. By
denoting the value of a given physical quantity with the same symbol as the
quantity itself (e.g., $x L\, \rightarrow \, x$ in the case of coordinate
$x$), the Schr\"odinger equation (\ref{eq:tdse8}) can be recast
\begin{equation}
\frac{\partial\Psi}{\partial\tau} \;=\; \frac{\hbar T}{2mL^2}
\frac{\partial^2\Psi}{\partial x^2} - \frac{T {\cal E}}{\hbar}
\left[V(x)-E_R\right]\Psi\;.
  \label{eq:du1}
\end{equation}
It is convenient to choose $L$, $T$, and ${\cal E}$ such that

\begin{equation}
\frac{\hbar T}{2mL^2} \;=\; \frac{1}{2}\;, \qquad \mbox{and} 
\qquad \frac{T {\cal E}}{\hbar} \;=\; 1
  \label{eq:du2}
\end{equation}
holds. 
Since there are three unknown units and only two relationships between
them, one has the freedom to specify the actual value of either 
$L$, $T$, ${\cal E}$ while the value of the other two units follows from
(\ref{eq:du2}).

In dimensionless units obeying (\ref{eq:du2}) the original
imaginary--time Schr\"odinger equation reads
\begin{equation}
\frac{\partial\Psi}{\partial\tau} \;=\; 
\frac{1}{2}\frac{\partial^2\Psi}{\partial x^2} - 
\left[V(x)-E_R\right]\Psi\;.
  \label{eq:du3}
\end{equation}

It is not difficult to transcribe also the other relevant equations
above in dimensionless units; for example, the functions
(\ref{eq:pif4}, \ref{eq:pif5}) become
\[ 
P\left(x_n,x_{n-1}\right) \;=\; \sqrt{\frac{1}{2\pi\Delta\tau}}
\exp\left[ -\frac{\left(x_n-x_{n-1}\right)^2}{2\Delta\tau}\right]\;,
\]
and 
\[
W\left(x_n\right) \;=\;\exp\left\{-\left[V\left(x_n\right)-E_R\right]\right\}\;,
\]
respectively. As a consequence, the diffusive displacements (\ref{eq:pif8a}) are described by
\begin{equation}
x_n\; = \; x_{n-1}\, + \, \sqrt{\Delta\tau} \rho_n
  \label{eq:du3a}
\end{equation}

\subsection{Computer Program}
\label{sec:cprog}

\begin{figure}[tb]
 \centerline{\epsfxsize=3in\epsffile{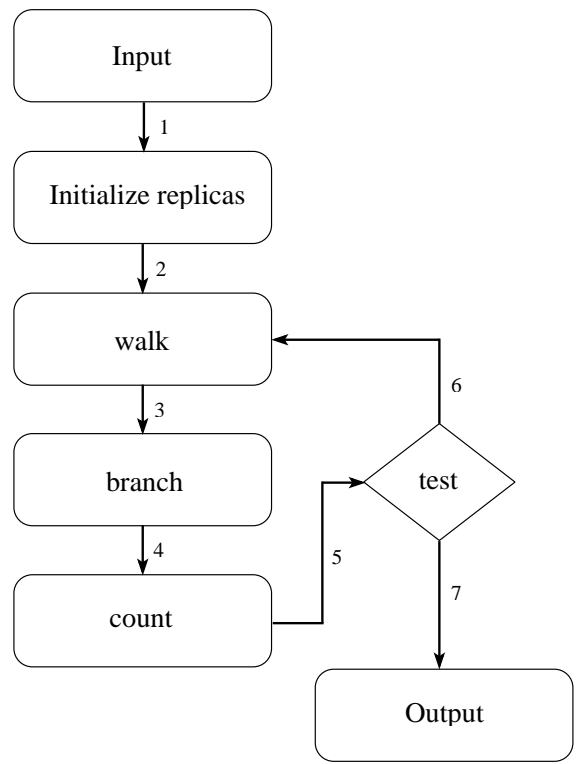}}
 \caption{Flow diagram of the DMC algorithm.}
 \label{fig:flow}
\end{figure}

The flow diagram of the computer program\cite{note:prog} implementing
the DMC method is shown in Fig.~\ref{fig:flow}. Each
block in the diagram performs specific tasks which are explained now.

All external data required for the calculation are collected through a
menu driven, interactive interface in the {\tt Input} block.  First,
one has to select the quantum system on which the calculation is
performed. At the program level this means to define the right spatial
dimensionality $d$ and the potential energy $V$ (see
Sec.~\ref{sec:app}) which corresponds to the selected system. The
quantum systems covered by our program are the ground state of the
harmonic oscillator, of the Morse oscillator, of the hydrogen atom, of
the H$_2^+$ ion (electronic state) and of the hydrogen molecule
(electronic state). The results of the simulation for all these cases
are presented in Sec.~\ref{sec:app} below. The other input parameters
are: the initial number of replicas $\left({\cal N}_0\right)$, the
maximum number of replicas $\left({\cal N}_{max}\right)$, the {\tt
  seed} value for the random number generators, the number of time
steps to run the simulation $(\tau_0)$, the value of the time step
$(\Delta\tau)$, the limits of the coordinates for the spatial sampling
of the replicas $\left(x_{min},\: x_{max}\right)$ and, finally, the
number of spatial ``boxes'' $\left( n_b\right)$ for sorting the
replicas during their sampling.  Suggested values for these parameters
are (in dimensionless units): ${\cal N}_0 = 500$, ${\cal N}_{max} =
2000$, $\tau_0 = 1000$, $\Delta\tau = 0.1$, $x_{min} = -20$, $x_{max} =
20$ and $n_b = 200$.

In the next block, {\tt Initialize replicas}, a two dimensional
matrix, called {\tt psips}, is defined with the following structure:
The first (row) index identifies the replicas and takes integer
values between one and ${\cal N}_{max}$. The second (column) index
points to information regarding the replica identified by the first
index, and is a positive integer less or equal to the number of
degrees of freedom of the system (i.e., $d'$, see Sec.\ref{sec:sdf})
plus one. For a given replica, say $i$, {\tt psips[i][1]} is used as
an {\em existence flag\/}: it is zero (one) if the replica is {\em
  dead\/} ({\em alive\/}). The other elements {\tt psips[i][j]}, $(j =
2,\ldots,d'+1)$ are used to store the coordinates of the replica
(i.e., $x_{j,\alpha}$ according to the notations of Sec.\ref{sec:sdf})
during the simulation.

To initialize the matrix {\tt psips} one sets equal to one the value
of the existence flag {\tt psips[j][1]} for $j = 1,\ldots {\cal N}_0$
( replicas $j \, \ge \, {\cal N}_o$ are not born yet and, accordingly, their existence flag
is set to zero), and then one assigns the same coordinates for all these
replicas (for the actual values of these coordinates,  see
Sec.\ref{sec:app}). Such a choice for the initial distribution of the
replicas corresponds to a $\delta$--function for $\Psi(x,0)$.
For a suitable choice of $x_0$ one  can be certain that there is always a significant overlap of 
$\Psi(x,0)$ and the ground state wave function $\phi_0(x)$. The
initial value of the reference energy $E_R$ is simply given by the
value of the potential energy at the initial position of the replicas
[c.f.\ (\ref{eq:dmcm4b})].

After initializing the replicas the program enters into  a loop which
essentially consists of three routines: {\tt walk}, {\tt branch} and
{\tt count}. One loop corresponds to taking one time step $\Delta\tau$.

\noindent {\tt walk}:
This routine performs the diffusion process of the replicas by adding to the
coordinates of the active (alive) replicas the value $\sqrt{\Delta\tau}\,
\rho$, where $\rho$ is a random number drawn from a Gaussian distribution
with mean zero and standard deviation one [c.f.\ (\ref{eq:du3a})]. The
program uses for this purpose the Gaussian random number generator {\tt
  gasdev}\cite{nr-C} .

\noindent {\tt branch}:
The birth--death (branching) processes, which follow the diffusion steps of
the replicas, are performed by the routine {\tt branch}.  For each alive
replica the number $m_n$, given by (\ref{eq:pif9af}), is calculated.  For
the generation of $u$ in (\ref{eq:pif9af}), a uniformly distributed random
number in the interval $[0,1]$, the program employs the function {\tt
  ran3}\cite{nr-C}. If a value $m_n = 0$ results, the replica is killed by
setting the corresponding existence flag to zero. If a value $m_n =1$
results, the replica is left as it is. If a value $n = 2$ results, the
replica is duplicated (the first inactive replica in {\tt psips} is set
active with the same coordinates as the original replica). If a value $n\, =
\, 3$ results, two identical copies of the replica are generated in {\tt
  psips}.  The reader may note that never more than two copies are born;
this limitation is necessary to prevent the uncontrolled growth of the
number of replicas which {\tt psips}, due to its finite size, might not be
able to accommodate. Such growth might occur when all the replicas are
located in (almost) the same place and when one can expect that large values
$m_n$ can arise. Since (by choice) the replicas initially are located in one
and the same point of the configuration space the first diffusion process
does not spread the replicas far enough and, for certain initial positions,
$m_n$ could then become large for all replicas. The algorithm also
terminates if all the $m_n$'s assume values zero. To avoid this possibility
one needs to choose the initial location of the replicas with care. In
general, any point where the ground state wave function is large is a good
choice.

\noindent  {\tt count}:   The role of this routine is to return the ground state wave
function of the system (i.e., the spatial distribution of the replicas) at
the end of the simulation. To this end, the spatial interval
$\left(x_{min},x_{max}\right)$ is divided equally into $n_b$ ``boxes''
(sub-intervals) for each degree of freedom and then, by employing standard
numerical methods\cite{koonin}, one counts the distribution of replicas
among these ``boxes''. The counting process starts after $\tau_0$ (in units
of $\Delta \tau$) time steps when the system has already reached its
stationary state (identified through a converged $\langle V\rangle_\tau$)
and is performed in a cumulative way for another $\tau_0$ time steps. This
strategy can be justified as follows: once stationarity is reached, the wave
function $\Psi(x,\tau) \propto \phi_0(x)$ will practically not change in
time; hence, the replicas will essentially sample one and the same wave
function at any subsequent time and the cumulative counting of the replicas
in the ``boxes'' can be used; this procedure yields better statistics for
$\phi_0(x)$ than sampling of replicas at only one instant in time; by
cumulative counting, the effective number of replicas used to sample the
wave function is enhanced by a factor of $\tau_0$ (number of time steps the
counting is done).

Once the spatial distribution of replicas is known, one can normalize
the distribution to obtain the ground state wave function using 
\begin{equation}
  \phi_0(x_i) \approx \frac{N_i}{\sqrt{\sum_{i=1}^{n_b}N_i^2}}\;, \quad
  i = 1,\ldots,n_b\;.
  \label{eq:cprog3}
\end{equation}
As should be apparent by now, until the completion of the algorithm
the routines {\tt walk} and {\tt branch} are called $2\tau_0$ times
while the {\tt count} routine is called only $\tau_0$ times, namely,
during the second half of the calculation. This is controlled by the
block {\tt test} shown in Fig.~\ref{fig:flow}.

Finally, the {\tt Output} block returns the results of the simulation.
These results are: (i) the average value of the reference energy $\langle
E_R\rangle \approx E_0$, which is calculated during the second part of the
simulation (i.e., from $\tau_0$ to $2\tau_0$) when the system is already
stabilized; (ii) the corresponding standard deviation $\delta E_R$; (iii)
the (imaginary) time evolution of $\langle E_R\rangle$ for the first
$\tau_0$ time steps (used basically to check how fast stationarity is
reached by the system during the simulation); (iv) the normalized spatial
distribution of the replicas, i.e., the ground state wave function. Note
that the average reference energy after $n$ time steps is defined through
\begin{equation}
  \label{eq:cprog4}
  \langle E_R(\tau=n\Delta\tau)\rangle \;=\; \frac{1}{n}
  \sum_{i=1}^n E_R(i\Delta\tau)\;.
\end{equation}

\section{Examples}
\label{sec:app}

In this section we report on the results obtained,  by means of the DMC program,  for the ground state energy and wave function of some quantum mechanical systems. The program
was executed on an  HP-9000 (series 700) workstation. In each case we
specify the units used and present numerical
results of the simulation. For all simulations the  values of
the input parameters suggested in Sec.~\ref{sec:cprog},  have been employed.

\subsection{Harmonic Oscillator}
\label{sec:hosc}

For a one--dimensional harmonic oscillator, characterized by the proper
angular frequency $\omega$, one has (in SI units)
\begin{equation}
V(x) \;=\; \frac{1}{2}m\omega^2 x^2\;.
  \label{eq:hosc1}
\end{equation}
Choosing $T = 1/\omega$, one obtains from (\ref{eq:du2}) $L
= \sqrt{\hbar/m\omega}$ and ${\cal E} = \hbar\omega$. 
In corresponding  dimensionless units, the exact ground state energy is $E_0 = 1/2$
and the  ground state  wave function is\cite{landau3} 
\begin{equation}
\phi_0(x) \;=\; \pi^{-\frac{1}{4}} \exp\left(-\frac{x^2}{2}\right)\;.
  \label{eq:hosc2}
\end{equation}

\begin{figure}[tb]
 \centerline{\epsfxsize=3.2in\epsffile{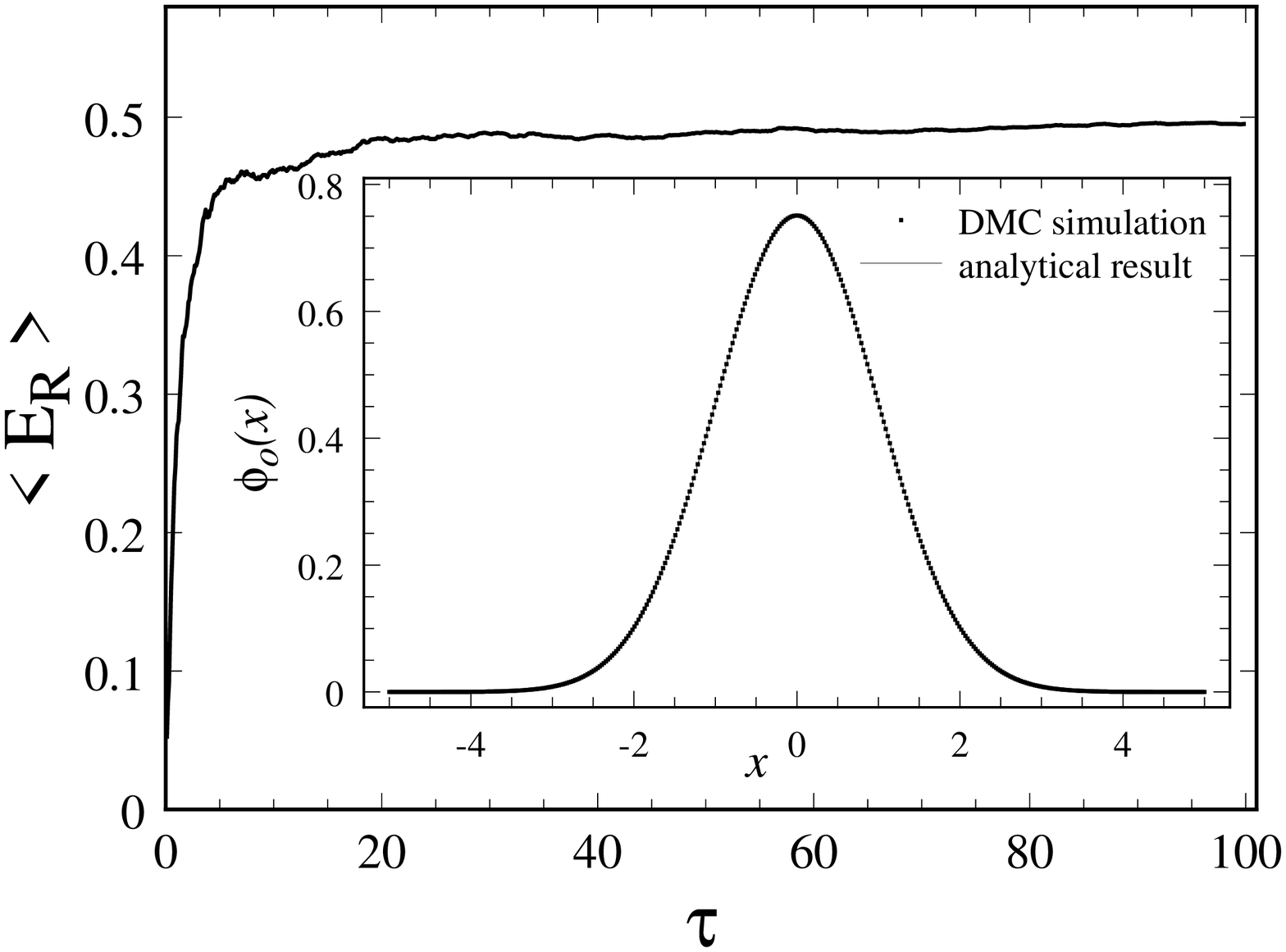}}
 \caption{Reference energy $E_R$ (in dimensionless units) as a
   function of the imaginary time $\tau$ (in units of time steps
   $\Delta\tau$) obtained by the DMC method for the
   harmonic oscillator. $E_R$ converges rapidly towards the exact
   ground state energy $E_0 = 0.5$. The inset shows the corresponding
   ground state wave function $\phi_0(x)$.  The result of the
   simulation is represented by dots while the continuous line
   corresponds to the analytical solution (\protect\ref{eq:hosc2}) }
\label{fig:hosc1}
\end{figure}

The results of the Monte Carlo simulation for the harmonic oscillator are
given in Fig.~\ref{fig:hosc1}. At the beginning of the simulation all
replicas are located at the origin. The main graph shows the rapid
convergence of the reference energy $\langle E_R(\tau)\rangle$ towards the
exact ground state energy $E_0$. The inset contains the plot of the ground
state wave function: the results of the simulation are represented by
triangles, while the continuous line corresponds to the analytical result
(\ref{eq:hosc2}). The agreement between the diffusion Monte Carlo result and
the analytical expressions is very good.

\subsection{Morse Oscillator}
\label{sec:morse}

The Morse potential is defined through

\begin{equation}
V(x) \;=\; D\left(e^{-2ax}-2e^{-ax}\right)\;.
  \label{eq:morse1}
\end{equation}
In this case one has a natural length scale through $L = 1/a$ which can
be used as the unit of length. As a result, from
(\ref{eq:du2}), one has $T = m/\hbar a^2$ and ${\cal E} =
\hbar^2 a^2/m$. For simplicity we shall consider only the
case when (in dimensionless units) $D = 1/2$; the exact
ground state energy is $E_0 = -1/8$, and the corresponding wave
function is\cite{infeld}
\begin{equation}
\phi_0(x) \;=\; \sqrt{2} \exp\left(-e^{-x}-\frac{x}{2}\right)\;.
  \label{eq:morse2}
\end{equation}

\begin{figure}[tb]
   \centerline{\epsfxsize=3.2in\epsffile{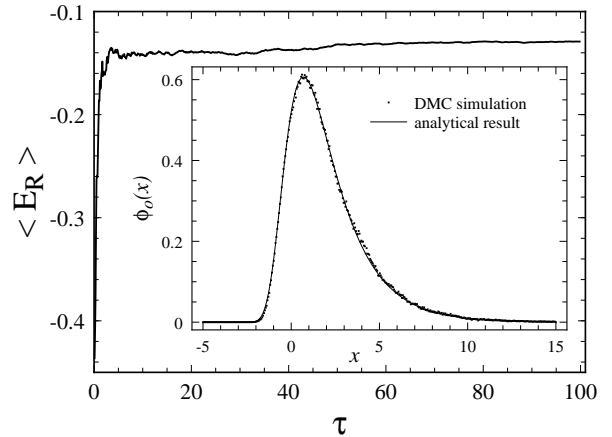}} 
   \caption{DMC description of the Morse
   oscillator. The time evolution of $E_R$ is shown.  The inset
   presents the ground state wave function; the numerical result is
   represented through dots, the wave function in the analytical
   form (\protect\ref{eq:morse2}) is represented by a continuous
   line.}
 \label{fig:morse1}
 \end{figure}

The results of the DMC description are presented in Fig.~\ref{fig:morse1}.
Initially, all replicas were positioned in our simulation at the origin. The
figure represents the time evolution of $\langle E_R\rangle$ towards the
exact ground state energy $E_0 = -0.125$.  The figure demonstrates also that
the the resulting ground state wave function is in excellent agreement with
(\ref{eq:morse2}).

\subsection{Hydrogen Atom}
\label{sec:hatom}

In case of the hydrogen atom it is customary to choose the unit of
length equal to the Bohr radius $a = \hbar^2/me^2 (\approx
0.53\AA)$. Thus, by setting $L = a$, one finds $T =
\hbar^3/me^4$ and ${\cal E} = me^4/\hbar^2 (\approx
27.21 eV)$. The well--known ground state energy (in dimensionless
units) is  $E_0 = -1/2$ and the corresponding radial wave
function is\cite{landau3} 
\begin{equation}
\phi_0(r) \;=\; 2e^{-r}\;.
  \label{eq:hatom1}
\end{equation}

\begin{figure}[tb]
 \centerline{\epsfxsize=3.2in\epsffile{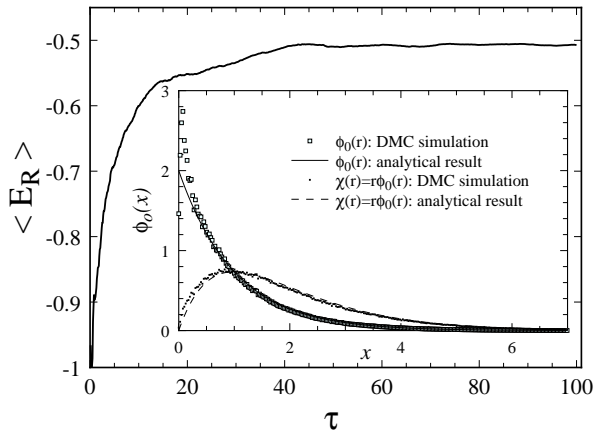}}
 \caption{DMC description of the hydrogen
   atom. The time evolution of $E_R$ is shown. The inset presents both
   the radial wave function $\phi_0(r)$ and the function
   $\chi(r) = r\phi_0(r)$.}
 \label{fig:hatom1}
 \end{figure}

We have carried out a DMC simulation for the hydrogen atom, generalizing the
previous description to three spatial dimensions.  At $\tau = 0$ all the
replicas were located at $(0,0,1)$, i.e., at a distance of one Bohr radius
from the origin, along the positive $z$ axis. Figure~\ref{fig:hatom1} shows
the convergence of $\langle E_R\rangle$ to the exact ground state energy.
This convergence, however, is not as rapid as in the case of the harmonic
oscillator and the Morse oscillator. This is not surprising since the
hydrogen atom has three degrees of freedom which require more sampling. The
running time of the simulation has also increased (see Table~\ref{tab:1}).
The inset shows both the radial wave function $\phi_0(r)$ (triangles) and
the function $\chi(r) = r\phi_0(r)$ (squares). For comparison, the
corresponding analytical solutions are also plotted with continuous lines.
Again the agreement between the analytical results and those obtained by our
algorithm are good.  The error of the radial wave function in the vicinity
of the origin is due to insufficient sampling of the number of replicas in
this region of the configuration space. To improve the wave function for $r
\sim 0$ one may decrease the size of the counting ``boxes'' in the vicinity
of the origin, increase the number of replicas, or increase $\tau$.

\subsection{H$_2^+$ Ion}
\label{sec:h2i}

\begin{figure}[tb]
 \centerline{\epsfxsize=3.2in\epsffile{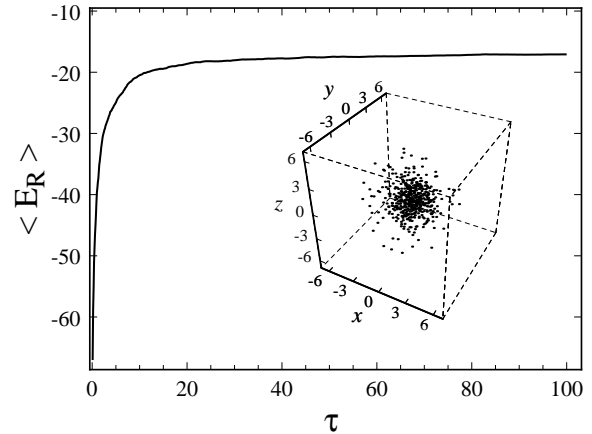}}
 \caption{DMC description of the H$_2^+$ ion.  
   The time evolution of $E_R$ is shown. The inset presents the
   spatial distribution of the replicas, i.e., the electron cloud.}
 \label{fig:h2i1}
 \end{figure}

The H$_2^+$ ion is held stabilized through a single electron moving in
the electric field of two protons separated, at equilibrium,  by a distance $R =
1.06\AA$\cite{herzberg}. The ground state of
this quantum system can not be determined analytically. The
numerically determined ground state energy of H$_2^+$ is $E_0 =
-16.252\pm 0.002$ eV\cite{herzberg}. The DMC method
can be applied in this case as for the hydrogen atom.  The Hamiltonian is
\begin{equation}
\hat H\; = \; -\,{\hbar^2\over 2m}\nabla^2\; - \; {e^2\over \vert \bbox{r}\, + \,{1\over 2}  \bbox{R}\vert} \; - \; {e^2\over \vert \bbox{r}\, - \, {1\over 2} \bbox{R}\vert}
\label{eq:hamh2+a}
\end{equation}
where $\bbox{R}$ denotes the separation between the two protons.  The
results of the DMC description are presented in Fig.~\ref{fig:h2i1}. The
same dimensionless units as in the case of the hydrogen atom were
employed and replicas were located initially at the origin, the nuclei
being located at $(0,0,\pm R/2)$, for $R = 2$. Figure~\ref{fig:h2i1}
shows that the  ground state energy obtained asymptotically  is $-16.75$ eV (see
Table~\ref{tab:1}) which differs from the more exact numerical value\cite{herzberg} by $0.5$ eV,
i.e., by about $3$\%. The inset in Fig.~\ref{fig:h2i1} shows a plot of
the spatial distribution of the replicas, demonstrating that the
electronic ground state wave function is nearly spherically symmetric.

\subsection{H$_2$ Molecule}
\label{sec:h2mol}

The H$_2$ molecule is formed by two protons,  at equilibrium separated by $R =
0.74\AA$\cite{herzberg}, and by two electrons.  In the Born approximation one considers the proton positions fixed and solves the corresponding stationary Schr\"odinger equation for the electrons. The wave function of the two electrons must be antisymmetric with respect to exchange of the electrons. In the ground state the electrons are in a singlet spin state 

\begin{equation}
\chi(1,2) =  \sqrt{1\over 2}\, \left[ \chi_{{1\over 2}, +{1\over 2}}(1)   \chi_{{1\over 2}, -{1\over 2}}(2)  -  \chi_{{1\over 2}, -{1\over 2}}(1) \chi_{{1\over 2}, +{1\over 2}}(2) \right]
\label{eq: h2_1}
\end{equation}
which is antisymmetric. Here $\chi_{\frac{1}{2},\pm\frac{1}{2}}(1,2)$
denotes the wave function of electron $1,2$ in the spin $\frac{1}{2}$ state
with magnetic quantum numbers $\pm\frac{1}{2}$. Accordingly, in the wave
function of the electronic ground state
\begin{equation}
\Psi_0\; = \; \Phi(\bbox{r}_1, \bbox{r}_2)\, \chi(1,2)
\label{eq: h2_2}
\end{equation}
the factor $\Phi(\bbox{r}_1, \bbox{r}_2)$, describing the spatial distribution of the electrons,  must be symmetric.  This factor can be determined through 
\begin{equation}
 \Phi(\bbox{r}_1, \bbox{r}_2)\; = \; \sqrt{1\over 2} \left(\, \phi(\bbox{r}_1, \bbox{r}_2) \; + \; 
\phi(\bbox{r}_2, \bbox{r}_1)\, \right)
\label{eq: h2_3}
\end{equation}
where $ \phi(\bbox{r}_1, \bbox{r}_2)$ is a solution of 
\begin{eqnarray}
& & \left[\, -{\hbar^2\over 2m}\left(\nabla_1^2 \, + \, \nabla_2^2\right) \; - {e^2\over \vert \bbox{r}_1 - {1\over 2}\bbox{R}\vert}\; - \;  {e^2\over \vert \bbox{r}_1 + {1\over 2}\bbox{R}\vert}\; 
\right.  \nonumber \\
& & \left. \qquad  
 - \;  {e^2\over \vert \bbox{r}_2 - {1\over 2}\bbox{R}\vert}\; - \;  {e^2\over \vert \bbox{r}_2 + {1\over 2}\bbox{R}\vert}\; + \; {e^2\over \vert \bbox{r}_1 - \bbox{r}_2\vert}\, \right]\, \phi(\bbox{r}_1, \bbox{r}_2)\; \nonumber \\[7pt]
& & \qquad \qquad \qquad \qquad \qquad \qquad\qquad = \;   E_0\, \phi(\bbox{r}_1, \bbox{r}_2)
\label{eq: h2_4}
\end{eqnarray}

\widetext
\begin{table}[d]
  \caption{Results obtained during two different simulations performed
    on four quantum systems listed in the first column. During Simulation I
    (II) the initial number of replicas was 500 (4000), the time step
    $\Delta\tau = 0.1$ (0.05), the length of simulation $\tau = 1000
    \Delta\tau$ ($2000 \Delta\tau$), the random number generator seed 1 and
    the number of boxes used to calculate the spatial distribution of the
    replicas 200 (400). $\langle E_R \rangle$ and $\delta E_R$ are defined
    in the text.  $\Delta t$ represents the actual running time of the
    simulation on a HP-9000 (series 700) workstation. For the hydrogen atom
    the energies are given both in dimensionless units and in eV (in
    parenthesis), respectively. For comparison, $E_0$ in the last column
    represents the exact value (analytical or obtained through an
    alternative numerical method) of the corresponding ground state energy.}
 \label{tab:1}
 \begin{tabular}{lccccccc}
   & \multicolumn{3}{c}{Simulation I} & \multicolumn{3}{c}{Simulation II}\\
   Quantum system & $\langle E_R \rangle$ & $\delta E_R$ & $\Delta t$
   & $\langle E_R \rangle$ & $\delta E_R$ & $\Delta t$ & $E_0$ \\
   \hline
   Harmonic oscillator & 0.505 & 0.094 & 8 sec & 0.500 & 0.048 & 4 min
   & 0.5\\
   Morse oscillator & -0.1236 & 0.0749 & 10 sec & -0.1245 & 0.0330 & 4
   min & -0.125\\
   Hydrogen atom & -0.495 & 0.080 & 18 sec & -0.505 & 0.040 & 5 min &
   0.5\\
   & (-13.477 eV) & (2.186 eV) & & (-13.752 eV) & (1.093 eV) & &
   (13.6 eV)\\
   H$_2^+$ ion & -16.753 eV & 2.869 eV & 40 sec & -16.476 eV & 1.389
   eV & 11 min & -16.25(2) eV$^a$\\
   H$_2$ molecule & -30.973 eV & 3.638 eV & 55 sec & -31.968 eV &
   1.754 eV & 16 min & -31.6(87) eV$^a$
 \end{tabular}
 $^a$ Numerical values calculated based upon the heat of dissociation
 given by Hertzberg\cite{herzberg}. 
 \end{table}
\narrowtext

\begin{figure}[tb]
 \centerline{\epsfxsize=3.2in\epsffile{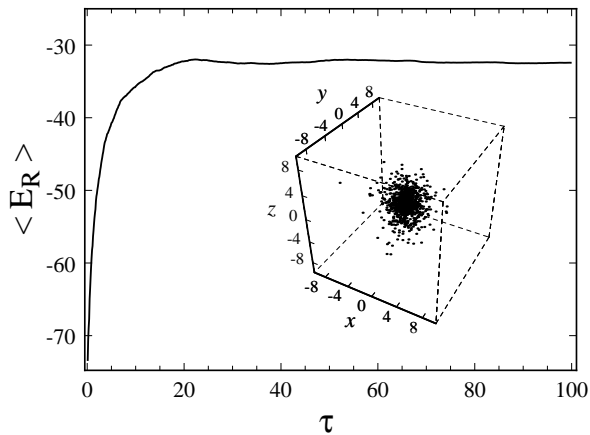}}
 \caption{DMC description of the electronic ground 
   state of the H$_2$ molecule.  The time evolution of $E_R$ is shown.
   The inset presents the spatial distribution of the replicas, i.e.,
   the electron cloud.}
 \label{fig:h2mol1}
 \end{figure}

\begin{figure}[tb]
 \centerline{\epsfxsize=3.2in\epsffile{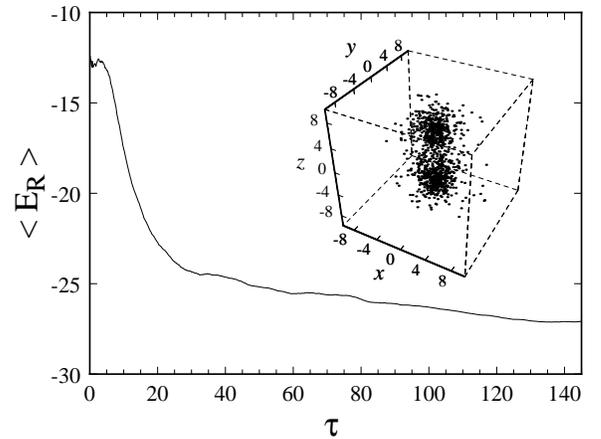}}
 \caption{$E_R$ vs.\ $\tau$ and the spatial distribution of the
   replicas (inset) at the end of the simulation for the H$_2$
   molecule for an unphysically large separation between the two
   protons of 8.398 Bohr radius.}
 \label{fig:h2mol2}
 \end{figure}

The Schr\"odinger equation (\ref{eq: h2_4})  cannot be solved analytically. The available
numerical result for the ground state energy of H$_2$ is $E_0 =
-31.688 \pm 0.013$~eV\cite{herzberg}. The results of the diffusion
Monte Carlo simulation for H$_2$ are shown in Fig.~\ref{fig:h2mol1}.
The same dimensionless units as in the case of the hydrogen atom were
employed and the replicas at $\tau = 0$ were located initially at
$\{(0,0,1)(0,0,-1)\}$; the position of the protons were $(0,0,\pm
R/2)$, with $R = 1.398$. The simulation took about $55$ seconds to complete and
the obtained ground state energy was $E_0 = -30.973$~eV (see
Table~\ref{tab:1}). This result differs from the exact energy by
$0.715$~eV, i.e., by $2.3$\%. The inset of Fig.~\ref{fig:h2mol1}  shows the
spatial distribution of the replicas which is nearly spherically symmetric. In
Fig.~\ref{fig:h2mol2} we present the results of a calculation in
which an  unphysically large  $R$ value of $8.398$ was assumed. In
this case the distribution of the electronic cloud around the protons is
clearly anisotropic, the energy of the electrons is still negative.

\section{Discussion}
\label{sec:con}

In this article, we have presented a detailed account of the path integral
formulation of the DMC method devised for calculating simultaneously the
ground state energy and wave function of an arbitrary quantum system. A
simple numerical algorithm based on the DMC method has been formulated and a
computer program, based on this algorithm, has been applied to determine
numerically the ground state of a few quantum systems of pedagogical
interest.

\begin{figure}[tb]
 \centerline{\epsfxsize=3.2in\epsffile{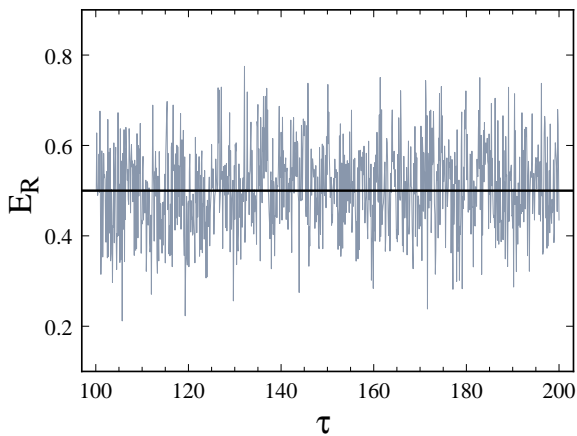}}
 \caption{Fluctuations of $E_R$ about its average value 
   $\langle E_R\rangle \approx 0.5$ well after the distribution of
   replicas reached its stationary form. The relatively large
   fluctuations are due to the birth--death processes which persist
   even if the replicas are distributed according to the exact ground
   state wave function $\phi_0(x)$ and the reference energy $E_R$ has
   been adapted to the exact ground state energy.}
 \label{fig:hosc2}
 \end{figure}

The DMC algorithm, as presented in this article, is quite unstable
numerically.  We want to demonstrate here the need for an improvement of the
method.  This need arises due to the strong fluctuations stemming from the
birth-death processes employed in this method.  These fluctuations affect
the reference energy $E_R(\tau)$ which is expected to converge to the ground
state energy.  This convergence is observed only for the value of $\langle
E_R\rangle$ defined through equation (\ref{eq:cprog4}). The actual time
dependence of $E_R$ is presented in Fig.~\ref{fig:hosc2} for the case of the
harmonic oscillator.  The fluctuations in $E_R$ are large, but symmetrically
distributed around $\langle E_R\rangle$ which explains the good agreement
between the result of the simulation with the exact result.  Attempts to use
our DMC program to calculate the electronic ground state energy of the
H$_2^+$ ion and H$_2$ molecule as a function of the separation between the
two protons, which would allow one to determine the equilibrium separation
from the minimum of this dependence, have failed due to large fluctuations
in $E_R$.

The DMC algorithm can be significantly improved resorting to a method called
``importance sampling''\cite{ceperley86,suhm91}. The basic idea of this
method is to change the probability distribution of the replicas in a
controlled way.  This can be achieved by reformulating (\ref{eq:tdse8}) such
that the resulting equation has a solution $\Psi(x,\tau)$ multiplied by an
approximation of the ground state wave function, the latter being obtained,
for example, from a variational method.  Application of the DMC method to
this new equation yields then replicas which spend more time in
``important'' regions of the configuration space where the wave function
$\Psi(x,\tau)$ is expected to be large.  For details regarding importance
sampling the reader is referred to the literature cited%
\cite{ceperley86,kalos86,suhm91}.

Any implementation of the DMC method leads to systematic errors due to the
use of a finite time step $\Delta\tau$. Apparently these errors can be
reduced to zero by choosing $\Delta\tau\rightarrow 0$. Unfortunately, this
is not the case. Even worse: making the time step shorter and shorter does
not only increase the needed computer time, but also renders the successive
generations of replicas more and more correlated such that their
distribution actually departs more and more from the ground state wave
function. For each quantum system investigated it is necessary to find the
most convenient value of $\Delta\tau$ which is short enough to produce small
systematic errors and at the same time is long enough to keep the successive
distributions of replicas sufficiently uncorrelated.

In order to apply the DMC method to calculate the
ground state properties of a system with interacting fermions one has
to treat, as mentioned above, the ``sign problem'' due to the
antisymmetry property of the many-fermion wave function. Two principle
methods, the {\em fixed--node\/} method\cite{reynolds82,ceperley86}
and the {\em release--node\/} method\cite{ceperley84,ceperley86}, have
been proposed to deal with this problem.

Once some kind of importance sampling is implemented and the
sign--problem, in the case of many--fermion systems, is resolved, the
DMC method can be used to compute ground state
properties for molecules or molecular clusters\cite{suhm91} and for
quantum spin, boson and fermion systems\cite{suzuki}. The method is
also applicable to the study of ground--state phase transitions due to
quantum fluctuations, a topic of modern condensed matter physics\cite{suzuki}.

Finally we would like to mention that the DMC method has been extended and
successfully applied to the study of the excited states of molecules and 
clusters\cite{suhm91} and also to the study of finite temperature properties
of different condensed matter systems. Also, the DMC method has been
successfully applied in quantum field theories\cite{barnes85}.

\section{Acknowledgments}

This work has been supported by a National Science Foundation REU fellowship
to B.F., by funds of the University of Illinois at Urbana-Champaign, and
by a grant from the {National Science Foundation (BIR-9318159)}.


\widetext

\begin{references}

\bibitem{ceperley86}
D.~Ceperley and B.~Alder, ``Quantum {M}onte {C}arlo'', {\em Science} {\bf 231},
  555--560 (1986).

\bibitem{kalos86}
M.~H. Kalos and P.~A. Whitlock, {\em Monte {C}arlo methods} (J. Wiley \& Sons,
  New York, 1986).

\bibitem{anderson}
J.~B. Anderson, ``A random--walk simulation of the {Schr\"odinger} equation:
  H$^+_3$'', {\em J. Chem. Phys.} {\bf 63}, 1499--1503 (1975).

\bibitem{nr-C}
W.~H. Press, S.~A. Teukolsky, W.~T. Vetterling and B.~P. Flannery, {\em
  Numerical recipes in C : the art of scientific computing} (Cambridge
  University Press, Cambridge, 1992), 2nd edn.

\bibitem{landau3}
L.~D. Landau and E.~M. Lifshitz, {\em Quantum Mechanics}, vol.~3 of {\em Course
  of Theoretical Physics} (Pergamon Press, Oxford, 1977).

\bibitem{feynman}
R.~P. Feynman and A.~R. Hibbs, {\em Quantum mechanics and path integrals}
 (McGraw-Hill, New York, 1965).

\bibitem{khandekar93}
D.~Khandekar, S.~Lawande and K.~B.~Hagwat, {\em Path Integral Methods and Their
  Applications} (World Scientific, London, 1993).

\bibitem{koonin}
S.~E. Koonin, {\em Computational Physics} (Benjamin, Reading, MA, 1986).

\bibitem{suhm91}
M.~A. Suhm and R.~O. Watts, ``Quantum Monte Carlo studies of vibrational states
  in molecules and clusters'', {\em Phys. Rep.} {\bf 204}, 293--329 (1991).

\bibitem{herzberg}
G.~Herzberg, {\em Molecular Spectra and Molecular Structure} (Van Nostrand, New
  York 1950).

\bibitem{note:prog}
The program was written is the C programming language. A copy of the source
  code is freely available from one of the authors (K.S.).

\bibitem{infeld}
L.~Infeld and T.~E. Hull, ``The factorization method'', {\em Rev. Mod. Phys.}
  {\bf 23}, 21--68 (1951).

\bibitem{reynolds82}
P.~J. Reynolds, D.~M. Ceperley, B.~J. Adler and J.~W.~A. Lester, ``Fixed--node
  quantum {M}onte {C}arlo for molecules'', {\em J. Chem. Phys.} {\bf 77},
  5593--5603 (1982).

\bibitem{ceperley84}
D.~M. Ceperley and B.~J. Adler, ``Quantum {M}onte {C}arlo for molecules:
  Green's function and nodal release'', {\em J. Chem. Phys.} {\bf 81},
  5833--5844 (1984).

\bibitem{suzuki}
M.~Suzuki, {\em Quantum Monte Carlo methods in condensed matter physics} (World
  Scientific, Singapore, 1993).

\bibitem{barnes85}
T.~Barnes and G.~J.~Daniell, ``Numerical {S}olution of {F}ield {T}heories
{U}sing {R}andom {W}alks'', {\em Nucl. Phys.} B{\bf 257}, 173--198 (1985).

\end{references}
\end{document}